\documentclass[conference]{IEEEtran}
\IEEEoverridecommandlockouts
\usepackage{cite}
\usepackage{amsmath,amssymb,amsfonts}
\usepackage{algorithmic}
\usepackage{graphicx}
\usepackage{textcomp}
\usepackage{xcolor}
\def\BibTeX{{\rm B\kern-.05em{\sc i\kern-.025em b}\kern-.08em
    T\kern-.1667em\lower.7ex\hbox{E}\kern-.125emX}}
\begin{document}

\title{DEMAND RESPONSE FOR RESIDENTIAL USES: A DATA ANALYTICS APPROACH*\\
	\thanks{* We would like to thank the Natural Sciences  and Engineering Research Council of Canada (NSERC) for their support through the Strategic Grants Program.}}

\author{
	\IEEEauthorblockN{Abdelkareem Jaradat, Hanan Lutfiyya, Anwar Haque}
	\IEEEauthorblockA{\textit{Department of Computer Science} \\
	\textit{The University of Western Ontario}\\
	London, Canada \\
	\{ajarada3, hlutfiyy, ahaque32\}@uwo.ca}
}

\maketitle

\begin{abstract}
In the Smart Grid environment, the advent of intelligent measuring devices facilitates monitoring appliance electricity consumption. This data can be used in applying Demand Response (DR) in residential houses through data analytics, and developing data mining techniques. In this research, we introduce a smart system foundation that is applied to user's disaggregated power consumption data. This system encourages the users to apply DR by changing their behaviour of using heavier operation modes to lighter modes, and by encouraging users to shift their usages to off-peak hours. First, we apply Cross Correlation (XCORR) to detect times of the occurrences when an appliance is being used. We then use The Dynamic Time Warping (DTW) \cite{ratanamahatana2004everything} algorithm to recognize the operation mode used.

\end{abstract}

%todo---------------------------------------------------------------------------------------------
\section{Introduction}

Demand Response (DR) is used to lower the demand on power systems by having consumers reduce or shift their power usage \cite{eia}. Demand response may be used to increase demand during periods of high supply and low demand. DR  is considered a more cost-effective option than building more power stations \cite{golz2017does}. 
Demand response is not a new concept, but in most countries it still plays a limited role \cite{iea}. One challenge is that although power consumption  for residential purposes accounts for 40\% or more of  overall  power consumption,  DR techniques have been ineffective \cite{NRC}.  One reason is that most of the programs for residential demand response focus on a single value of power consumption representing the total power consumption \cite{Pouresmaeil}.  However, overall residential power consumption is generated by a diverse set of power consuming devices that includes cooling and heating units, washing machines, dryers, refrigerators, lighting, etc. The usage of many of the individual appliances can be adjusted.  The diverse nature of devices that consume power suggests that a single value of power consumption representing the aggregation of the values of different devices that consume power is unsuitable. There has been relatively little work that considers disaggregated demand. 

With smart meters and sensors that can measure power consumption per appliance it has become much easier to monitor power consumption per device. In the work presented in this paper, we are particularly interested in being able to detect the start and end time of an appliance usage as well as the operation mode being used.    Each operation mode is characterized by its running time and different cycles that the appliance  can run in. Activating an appliance in a certain operation mode consumes energy differently than other modes. Therefore, DR could be applied based on the outcomes of this analysis by \textbf{first,} finding the activation times of appliances. The power consumption data is processed in order to find the turn on times for the appliance. We obtain the time when the high load starts. Consequently, a recommendation could be sent to the consumer if the detected time falls within the on-peak time advising the consumer to shift the load either before or after the current time to avoid higher energy prices. \textbf{Second,} recognizing the operation mode of each appliance activation. Most of modern appliances have the option to run in different operation modes. For example, a washing machine could be programmed with three or four different operation modes. Each of these modes runs the inner components of the washing machine differently. Also, every mode has its own timing and cycles activated in different power levels during the running time. By determining these cycles, a model could be formed for each operation mode. This has the potential to serve as the foundation for a smart recommendation system that applies DR to provide consumers with recommendations about their consumption for different appliances. These recommendations come into place after applying the detection of appliance activation and the recognition of operation modes in such way to encourage the residents to shift their usages to off-peak periods of the day, when power is cheaper. Also the recommendations could advise consumers to change their behaviour of using lighter operation modes with less consumption.

The rest of the paper is organized as the following: In section \ref{related} we discuss previous work that addresses event detection and classification. Section \ref{analysis} discusses the data analysis of the power consumption. In section \ref{detection} we present a detection algorithm. Section \ref{dtw} discusses the algorithm to classify appliances operation modes. Section \ref{simulation} discusses the process of simulating power consumption. Section \ref{results} discusses the results and section \ref{conc} concludes the paper.   
%The rest of the paper is organized as the following: In section \ref{related} we discuss different approaches in the literature to address event detection and classification in power consumption time series data. Section \ref{analysis} discusses the data analysis of the power consumption for different appliances and the different steps that the appliance runs though. In section \ref{detection} we present a detection algorithm to detect the activation time of appliances. Section \ref{dtw} discusses utilizing a similarity measure algorithm to classify operation modes of appliances.  Section \ref{simulation} discusses the process of simulating power consumption data based on the outcomes of the data analysis to test the proposed techniques. Section \ref{results} discusses the results and section \ref{conc} concludes the paper.
%todo---------------------------------------------------------------------------------------------
\section{Related Work}
\label{related}
The literature describes many techniques used to analyze the electricity consumption data so that end user applications could be built on top of these approaches. These techniques primarily concerned with event detection and event classification in time series data. Typically the event is the activation and deactivation of appliances during its operation with the power distribution over time. This serves in defining attributes to target customers for appliance specific application with DR.
\subsection{Event Detection}
Event detection algorithms are used to detect load profiles. A load profile is the power distribution in a form of cycles over a single run for an appliance. Event detection is concerned with finding transition states for loads/appliances from aggregated power consumption data. Each transition state is characterized by a sudden change in the power value, which indicates activating an appliance or a change in the running cycle. Based on the determined timing for the transitions, load profile are extracted separately. This is what is referred to as Non-Intrusive Load Monitoring (NILM) \cite{wang2018review}.

Matched Filters methods involve a known signal (template signal) to be correlated with an unknown signal to detect the occurrence of the template in the unknown signal.
Rueda et.al.\cite{rueda2018transient} proposed a matched filter detection approach by convolving the appliance consumption signal with a conjugated time-reversed version of a manually extracted template.
Baets et.al.\cite{de2016event} presented an event detection method that uses Cepstrum Analysis in the frequency domain.
Alcala et.al. \cite{alcala2014event} proposed an approach using Hilbert Transform to extract the envelope of the current signal. Then, by using Average Filters, Derivation Filters, and thresholding to cut off the signal, transition events are detected.

\subsection{Event Classification}
Different approaches are concerned with classifying load signature extracted from  aggregated power consumption data into the individual appliance/load. 
Liao et.al. \cite{liao2014power} proposed an approach for appliance load classification using Dynamic Time Warping (DTW).
Tang et.al. \cite{tang2015meter} designed the Sparse Human Action Recovery with Knowledge of appliances (SHARK) framework. It is an occupancy detection framework that is non-intrusive and requires no training process. 
Liu et.al. \cite{liu2017dynamic} used a Nearest Neighbor Transient Identification method to identify the appliance creating the Transient Power Waveform (TPW) sample time-series, then the DTW-based integrated distance is utilized to calculate the similarity of TPW signatures and a template time series for an appliance. 
Wang et.al. \cite{wang2018iterative} describes an approach which uses Iterative Disaggregation based on Appliance Consumption Pattern (ILDACP). This approach combines Fuzzy C-means clustering algorithm to detect appliance operating status, and DTW search that identifies single energy consumption based on the appliance typical power consumption pattern (a template pattern). 
\subsection{Gap Analysis}
Most of the work currently in the literature uses energy disaggregation (NILM) on aggregated power consumption data to determine if an appliance is activated. To the extent of our knowledge, the literature lacks approaches that focus on analyzing disaggregated power data and detect the activation of certain appliances and then classify each use of the appliance in one of its operation modes.
%todo---------------------------------------------------------------------------------------------
	\begin{figure}
	\centering
	\fbox{\includegraphics[trim={1.7cm 1.7cm 1.7cm 1.7cm },clip,width=.47\textwidth]{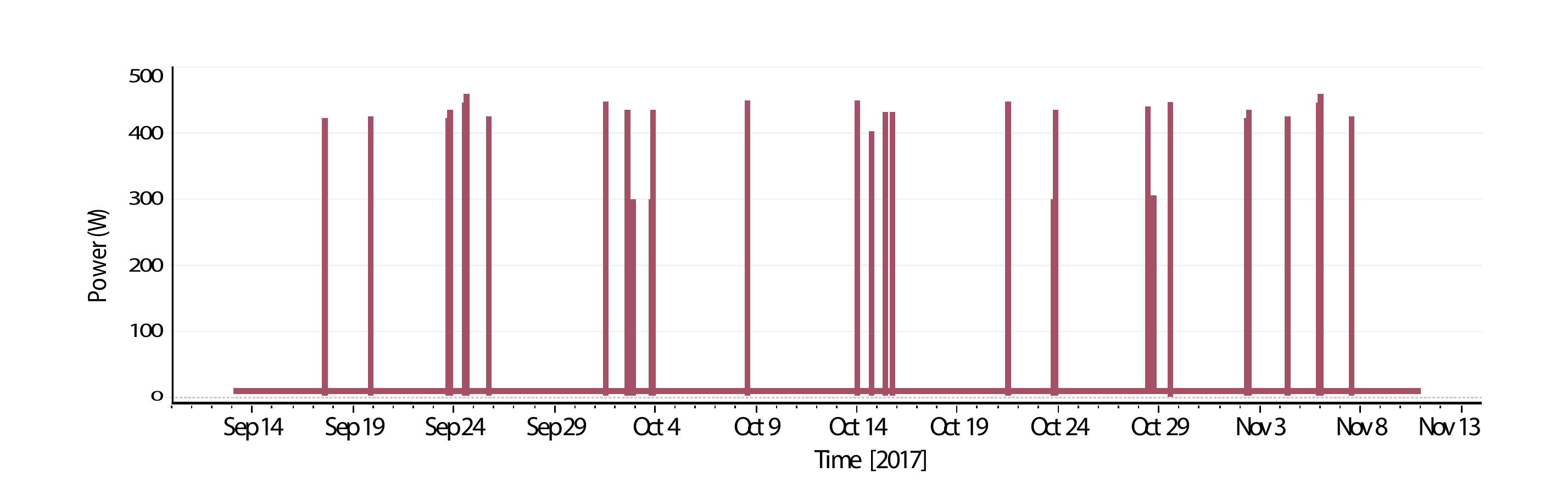}}
	\caption{Usage times for the clothes washer in House 2. Each spike shows a single operation for the washer.}
	\label{fig:Washwe_uses_powers_h2}
\end{figure}
\section{Power Consumption Data Analysis}
\label{analysis} 
We use Power Consumption Datasets (PCDs) to understand the power consumption characteristics of appliances over time when turned on. 
We focused on the analysis of The Rainforest Automation Energy dataset (RAE) \cite{Makonin2017} and the work done by Pipattanasomporn et.al. \cite{Pipattanasomporn2014} where demand response opportunities were provided for some household appliances. In this paper, the appliances data that we analyzed are the clothes washer, clothes dryer, and dishwasher. 
%=============================================
\subsection{Single Usage Profile}
A Single Usage Profile (SUP) is used to formally characterize power consumption of an appliance between the time it is turned on and the time it is turned off.  The analysis of the datasets is presented in \ref{analysisAppliances}  and is used to determine SUPs.
Single Usage Profile (SUP) represents the sequence of power consumption values consumed by an appliance from the moment of turning it on to the moment of that it is turned off. Hence, SUP with the sampling frequency $f_s=1Hz$  is defined by the sequence $\{ p_i \}_{i=t_{on}}^{t_{off}}$ where $p_i$ is the instantaneous power reading at time $i$. The times $t_{on}, t_{off}$ is the turn on, turn off times respectively of the appliance and $i \in [t_{on}, t_{off}]$ represents the $i^{th}$ sample. Figure \ref{fig:Washwe_uses_powers_h2} shows multiple SUPs for a clothes washer over the course of around two months.

%=============================================
\subsection{The Analysis Of Appliances Power Consumption}
\label{analysisAppliances} 
\begin{figure}
	\centering
	\fbox{\includegraphics[trim={1.7cm 1.7cm 1.7cm 1.7cm },clip,width=.47\textwidth]{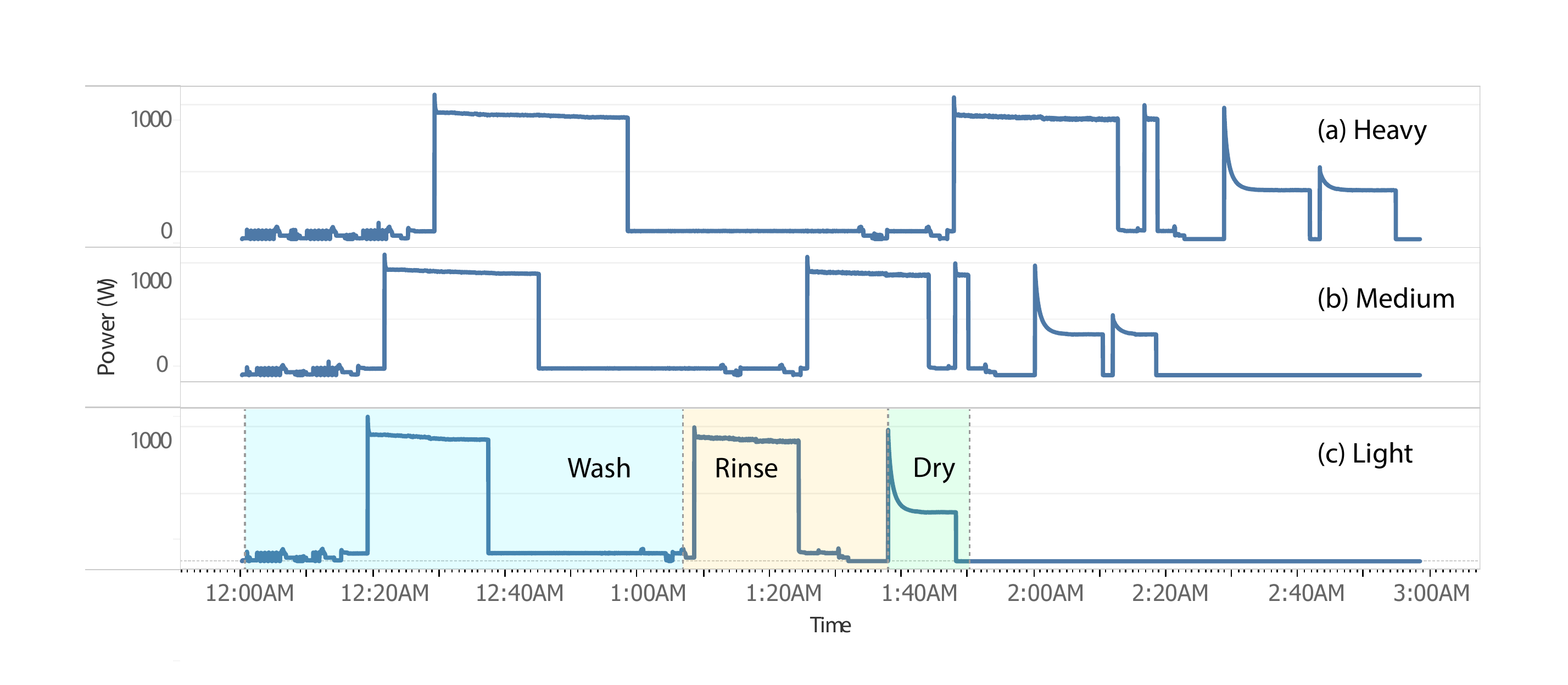}}
	\caption{Single Usage Profile (SUP) for the dishwasher showing three operation modes: (a) Heavy, (b) Medium, (c) Light.}
	\label{fig:Dishwasher_All_modes}
	\end{figure}

The SUP for the dishwasher is shown in Figure \ref{fig:Dishwasher_All_modes}. The dishwasher has three main operating modes: Heavy, Medium, and Light. Each SUP of the dishwasher has three states: Wash, Rinse and Dry, regardless of the operation mode. For example, in the Light mode, the first 70 minutes are associated with the wash state. The rinse state follows from minutes 70 to 97. The last state is the drying state which ends at minute 108.

The same type of analysis is performed for the clothes washer and the clothes dryer. The clothes washer has three states that occur for each single use: wash, rinse, and spin, sorted in ascending order of power consumption level.  A single operation of the clothes dryer uses three states with different temperatures: Maximum Heat, Medium Heat, No Heat. During the Maximum Heat state the dryer drum is heated to the maximum temperature in the dryer setting, and so on for the other states. 

By analyzing appliances data, we found similar patterns. Each activation with certain operation mode has a specific pattern of consumption based on the states of the appliance. These states are preconfigured by the manufacturer in terms of timing and power consumption levels.

%todo---------------------------------------------------------------------------------------------
\section{Single Use Profile Detection}
\label{detection} 
With Demand Response,  consumers are encouraged to use appliances in off peak hours. This requires the need to detect when appliances are activated. Hence, we can advice the user to shift the load or not based on the detected time. This serves as a basis for further analysis in determining appliance operation modes.

%=============================================
\subsection{The Reference Pattern}
We use a Reference Pattern (RP) which represents the start of a SUP. RP is a sequence of data points represented as $ S(t) $ which returns the data point that represents the power value associated with $ t $. We focus on matching RP to a subsequence of the daily time series which is represented by $ D(t) $. RP is derived from a generated SUP by taking a slice of the generated SUP corresponding to turning on the appliance.
%=============================================
\subsection{Cross Correlation}
In time series analysis, Cross Correlation (XCORR) \cite{MENKE2016187} is a measure of similarity of two series as a function of the displacement of one series relative to the other series. Assume two time series represented by $ f(t) $ where $ t \in [1,n]$ and $ g(t) $ where $  t \in [1,m] $ and $ n < m $. Assuming that the number of moving windows is finite, the XCORR function, $ X(t) $, for $ f(t) $ with $  g(t) $ is defined as follows:
   	\begin{equation} \label{eq:det02}
		X(t)= (f\star g)(t)\ = \sum _{k=1 }^{n }f(k)g(t+k)
	\end{equation}
To avoid calculations with large numbers, we use another option by using the absolute difference instead of multiplication. If we use the absolute difference, the  XCORR function $X(t)$ between $S(t)$ and $D(t)$  is the following: 
	\begin{equation} \label{eq:det04}
	X(t)= (S\star D)(t)\ = \sum _{k=1 }^{n }\Big| S(k) - D(t+k)\Big| \quad,\quad t\in[1,m-n]
	\end{equation}
\noindent where $ m $ represents the number of daily samples in $ D(t) $ and $ n $ is the number of samples used in $ S(t) $.

A normalization step takes place to invert $X(t)$ over the y-axis by subtracting all correlation values from the average of the maximum power value for both  $S(t)$ and $D(t)$. This normalization makes the similarity between the two functions relative to their maximum values. Therefore, the two functions are more similar when the normalized correlation value is closer to the average of maximum values found in the sequences represented by $ S(t) $ and $ D(t) $. The normalized cross correlation function $ X(t)  $ is the following:
	\begin{equation} \label{eq:det05}
	X(t) = \frac{Max(S(t)) + Max(D(t))}{2} - \frac{1}{n} \sum _{k=1 }^{n }\Big| S(k) - D(t+k)\Big| \quad,
	\end{equation}
	\[
		\quad t\in[1,m-n]
	\]
	\begin{figure}
	\centering
	\includegraphics[width=.47\textwidth]{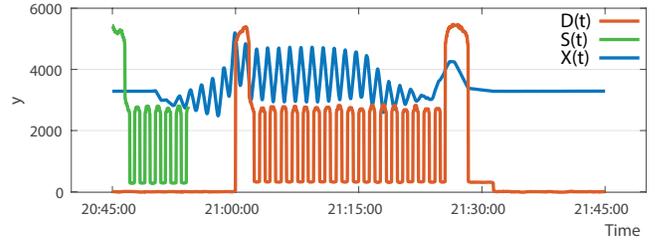}
	\caption{ The reference pattern represented by the function $ S(t) $. The cross correlation plot $ X(t) $ and the day consumption function $ D(t) $ over the day time. }
	\label{fig:detxcorr}
	\end{figure}
Figure.\ref{fig:detxcorr} shows a graphical depiction of the XCORR function $ X(t) = (S\star D)(t)$.
$ S(t) $ is shown in Figure.\ref{fig:detxcorr}. RP shows $ 600 $ samples in this case. $ X(t) $ is shown in Figure.\ref{fig:detxcorr} . The fluctuation in the value of $ X(t) $ starts around 8:50 PM as illustrated in Figure.\ref{fig:detxcorr} (b) which depicts a zoomed in view for the period 8:45 PM to 9:45 PM. These fluctuations represent the overlap between the two correlated functions. Higher values of $X(t)$  means that there is more overlap present between the functions and therefore, higher similarity. $X(t)$ shows its maximum value at approximately 9:00 PM, which means that the potential SUP starts at this time. The turn on time of the appliance in $D(t)$ is observed at 9:00 PM as seen in  Figure.\ref{fig:detxcorr} (b) representing the maximum value of $X(t)$.
%=============================================
\subsection{Determining Turn On Times}
 	\begin{figure}
	\centering
	\includegraphics[width=.47\textwidth]{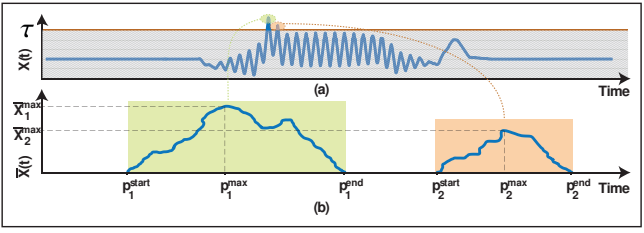}
	\caption{\textbf{(a)} The positioning of the threshold $\tau$ relative to the residue function $ \overline{X}(t) $. \textbf{(b)} A zoomed in version of $ \overline{X}(t) $. The plot shows the absolute maximum $X_i^{max}$ for each tip of spike  $i$ enclosed in the period $p_i=[ p_i^{start} , p_i^{end}]$ }
	\label{fig:detCancel}
\end{figure}
We use X(t) to determine the potential turn on times. It is not feasible to consider all times at the local maximums of $X(t)$ as potential turn on times since $D(t)$ could contain multiple periods where there is not a considerable level of similarity with the reference pattern. Also, the noise contained within the correlation function may cause local maximums even after a smoothing function is applied. To overcome these problems we introduce a Low Amplitude Canceling Coefficient $\delta$ that limits the range of where to look for the local maximums. The role of the $\delta$ coefficient is to cancel out all values of $ X(t) $ that is lower than a threshold $\tau$ determined by $\delta$, so that the residue function $\overline{X}(t)$ of $X(t)$  after applying $\delta$ represents all periods that contain the local maximums that is greater than $\tau$. The value of $\delta$ is selected with the assumption that $ \delta \in (0,1) $. $\overline{X}(t)$ is defined as the following:
	\begin{equation} \label{eq:det07}
	\overline{X}(t) = X(t) - \tau
	\end{equation}
\noindent where the value of the threshold $\tau$ is calculated as follows:
	\begin{equation} \label{eq:det06}
	\tau = \delta \;  \frac{Max(S(t)) + Max(D(t))}{2} \quad ,\quad \delta \in (0,1)
	\end{equation}
Figure.\ref{fig:detCancel} (b) shows the plot of  $ \overline{X}(t) $. Since $ \overline{X}(t) $ is a trimmed version of $ X(t) $ where  $ X(t) $ is thresholded by $\tau$ value to produce $ \overline{X}(t) $, then $ \overline{X}(t) $ is zero except for short periods of time. These periods of time are what remained from $ X(t) $ after applying the threshold $\tau$. $ \overline{X}(t) $ contains isolated periods of time where the value of the correlation is relatively the highest among the entire period of the day. The shape of the $ \overline{X}(t) $ is a group of continuous concave down curves that resembles the highest values of $ X(t) $ where $X(t) > \tau$. Therefore, the position of the potential turn on time would be at the absolute maximum of each of these concave down curves. These periods in Figure.\ref{fig:detCancel} are $p_1, p_2$ where $p_1=[ p_1^{start} , p_1^{end}]$ and $p_2=[ p_2^{start} , p_2^{end}]$ . These periods have their own absolute maximum values at $p_1^{max}, p_2^{max}$ respectively within their domains. This maximum value represents the highest value of the cross correlation at this period, which means that the time when this maximum values occur is a potential turn on time. Consequently, the absolute maximum in $p_1$ is $X_1^{max}$ at $t= p_1^{max}$, and the absolute maximum in $p_2$ is $X_2^{max}$ at $t= p_2^{max}$. We then conclude that $\{p_1^{max}, p_2^{max}\} $ is the set of the potential turn on times. 

%todo---------------------------------------------------------------------------------------------
\section{Single Use Profile Classification With DTW}
\label{dtw} 
For each time that the appliance has been activated we need to determine the appliance operation mode. This requires a comparison of the time series at the point that the appliance has been activated and the reference patterns for each of the operational modes. We describe the approach using Dynamic Time Warping (DTW) \cite{ratanamahatana2004everything} .

%=============================================
\subsection{Day Consumption Segmentation}
Let us assume that  an appliance runs in $ n $ operation modes denoted by the set $  M = \{m_1,m_2,...,m_n\} $.  The daily consumption is represented by $ D(t) $ which  contains subsequences  representing appliance usages.  For each mode $ m_i $  the reference pattern is represented by $ P^{m_i}(t) $  where $k^{m_i}$ is the SUP size of operation mode $ m_i $. A segment $ G^{m_i}(t)$ is a sub-sequence of the daily consumption function $ D(t) $, starting from the point of the turn on time $t_{on}$ and with the size $k^{m_i}$ representing the size of operation mode $m_i$. This is defined as follows:
\begin{equation} \label{eq:dtw03}
G^{m_i}(t) = D(x) \quad : \quad x \in [t_{on} , t_{on}+k^{m_i}]
\end{equation}

\noindent for each reference pattern, $ P^{m_i}(t)$, for an operation mode $ m_i $ and turn on time $t_{on}$, segments of the daily consumption, D(t) can be extracted using Eq.\ref{eq:dtw03}.

\begin{figure}
	\centering
	\includegraphics[width=.47\textwidth]{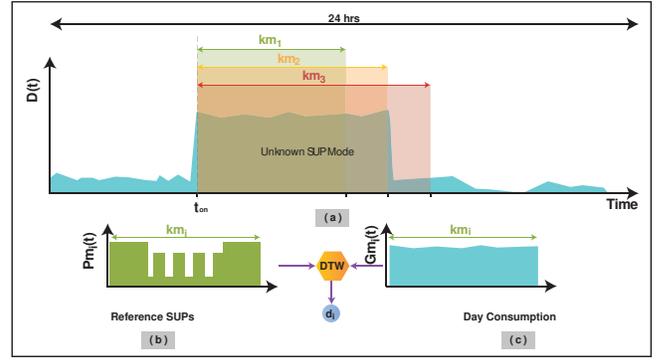}
	\caption{ Visual representation of the segmentation and applying DTW.}
	\label{fig:dtwSegmenter}
	\end{figure}

The segmentation is visually presented in Figure.\ref{fig:dtwSegmenter}. In part (a) $ D(t) $ is shown, and the segmentation starting point is indicated by the vertical dashed line at $t_{on}$. The segment sizes are highlighted in different shades so that each shade corresponds to the size of the segment for a specific operation mode. Part (b) shows the generated reference functions   $ P^{m_i}(t) $ that are used to specify segments sizes $ \{k^{m_1},k^{m_2},..,k^{m_n}\} $. Lastly, in part (c) the segment functions $ G^{m_i}(t)$ are listed.
%=============================================
\subsection{Classification of SUPs Based On DTW Distances}
The DTW algorithm \cite{ratanamahatana2004everything} uses two inputs which are the two functions $ G^{m_i}(t)$ and $ P^{m_i}(t) $ representing sequences. It then performs a calculation to find the distance (or similarity) between these two functions as the output. Figure.\ref{fig:dtwSegmenter} shows the inputs and the outputs of the DTW algorithm. The DTW is invoked $ n $ times where $ n $ equals the number of RPs representing operation modes for an appliance. For each reference pattern and the corresponding segment, the distance is calculated. Let X denote the sequence represented by $ P^{m_i}(t) $ and let Y represent the sequence  $ G^{m_i}(t)$.  The DTW distance is formulated as follows:

\begin{equation} \label{eq:dtw05}
\begin{array}{l}
d_i = \mathrm{DTW}(X,Y )
\\
M=\{m_1,..,m_i, ... m_n\}
\end{array}
\end{equation}

The value of the distance calculated by DTW is inversely proportional to the similarity.  Thus we assume that the most similar reference pattern is the reference pattern with the minimum distance. This means that the mode $m^*$ with the minimum distance $d^*$ that is the operation mode of the appliance at the SUP detected at $t_{on}$ as:
\begin{equation*} \label{eq:dtw06}
\begin{array}{l}
d^* = min (d_1,	d_2,	...,	d_{m_i})
\end{array}
\end{equation*}
%todo---------------------------------------------------------------------------------------------
\section{Power Consumption Simulation}
\label{simulation} 
It is necessary to create simulated data when it is impractical to obtain real data that there is an insufficient amount of certain type data such as our case when there is no such PCD that contains disaggregated data with different operation modes for an appliance. Our Power Consumption Simulator (PCS) main purpose is to simulate appliances consumption with different operation modes by generating daily usage data that has SUPs for different operation modes.

PCS has three configuration parameters: Turn On Time $t_{on}$, Household Usage Intensity ($I_h$), and SUP Representation Object (SUPRO). $t_{on}$ refers to the time when an appliance, \textit{a}, is activated by the user during the day. We use the Inverse Transform Sampling (ITS) \cite{miller2010inverse} method to sample values of $t_{on}$ Probability Density Function (PDF) that is extracted from a PCD. Household Usage Intensity ($I^a_h$) refers to the distribution of operation modes that a household, \textit{h}, uses for an appliance, \textit{a}, over time.  We assume that the selection of an appliance operation mode for a household is based on a multinomial distribution function. The frequencies of this distribution is 20\% for operation modes used fewer number of times and 60\% for the operation modes used more. Single Usage Profile Re presentation Object (SUPRO)  is a representational model of a SUP for an appliance in a particular operation mode.  It defines \textit{Cycles} which are periods of time when the power consumption is stable around fixed wattage. Each cycle is defined by the duration and the wattage it has. Also, SUPRO defines \textit{Phases} which are groups of cycles in certain repetition bounded by lower and upper bounds.

%todo---------------------------------------------------------------------------------------------
\section{Results And Discussion}
\label{results}
\begin{figure}
	\centering
	\includegraphics[width=.48\textwidth]{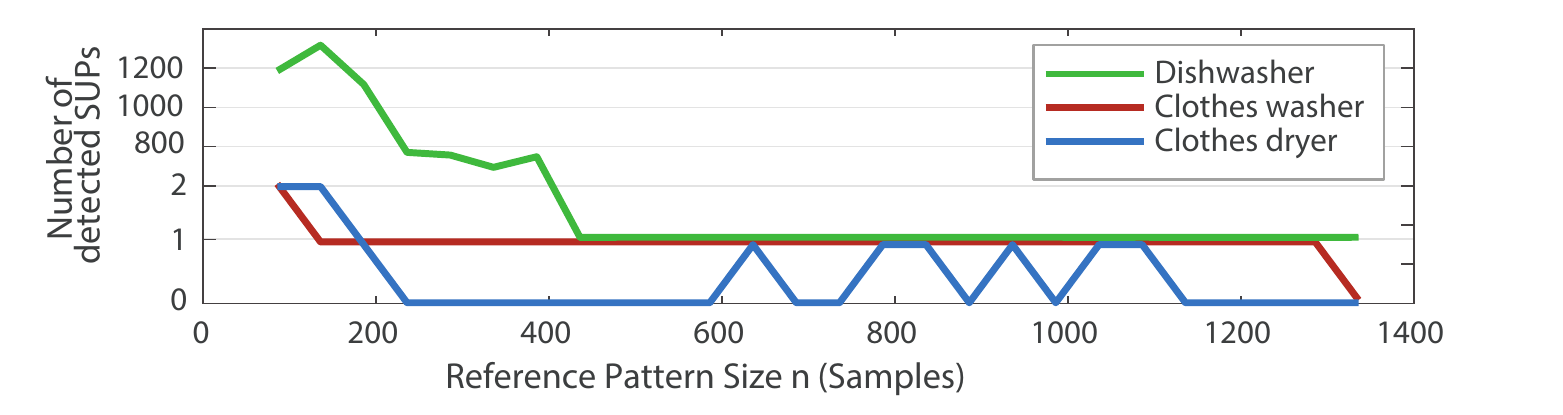}
	\caption{The impact of changing the reference pattern size $n$ on the number of detected SUPs in the day consumption date for a dishwasher, a clothes washer, and a clothes dryer. }
	\label{fig:allNumSUP}
\end{figure}
This section discusses the results of applying SUP detection technique using XCORR and SUP classification using DTW on a test data.
\subsection{Single Usage Profile Detection}
The evaluation metric we use is \textit{the number of detected SUPs } for certain appliance. This corresponds to the number of reported turn on times for certain appliance by the SUP detection algorithm.  We assume that a single SUP presents daily for an appliance. We generated reference patterns of different sizes $n$ based on a SUP generated with a randomly selected operation mode  (see \ref{simulation}).  We assume that the value of the Low Amplitude Canceling Coefficient $\delta$ is equal to 90 and the sampling frequency used is $f_s = 1Hz$.

Figure.\ref{fig:allNumSUP} shows the results for a dishwasher, a clothes washer, and a clothes dryer. For the dishwasher, it shows a very large number of detected SUPs found in the day when n is between 50 and 400. This is because of the shape of the SUP of the dishwasher. When the reference pattern size is  between 50 and 400, there is a high number of SUPs detected despite the existence of only one SUP. However, the number of detected SUP settles down to one when the reference pattern size typically increases within the range 400 and 1400 samples. When the reference pattern size exceeds 1400 samples, the number of detected SUPs is zero.
For the clothes washer. When the reference pattern size is between 100 and 1100 samples, the number of detected potential SUPs is always one. That is probably because the repetition in the washer consumption curve is minimal, thus, a unique one high value of cross correlation in every test.
For the clothes dryer, it shows that the value of $n$ that gives best results is when $n$ is between 600 and 1100 samples where it shows the number of detected SUPs equals to one. Nevertheless, the results show some choppiness in the number of detected SUPs. This is a common problem for all appliances. It is most likely due to the variations in cycles duration where cycle duration is generated with variation factor that affects XCORR by reducing the overlapping between the cycles in the reference pattern and the detected SUP.

The number of matching SUPs is higher for shorter reference pattern than a longer reference pattern. This is because there is higher chance that small reference pattern occurs more frequently within the consumption function than a longer reference pattern. The reason is that a SUP has repetitions of cycles that have similar shape. This shape could be similar to the reference pattern. This leads to reporting multiple potential SUPs even where is only one SUP. Thus, as the size of reference pattern decreases, the probability of reporting these repetitions within a SUP is higher. Therefore, the accuracy increases as the reference pattern size increases. 

\subsection{Single Usage Profile Classification}
	\begin{figure}
		\centering
		\includegraphics[width=.48\textwidth]{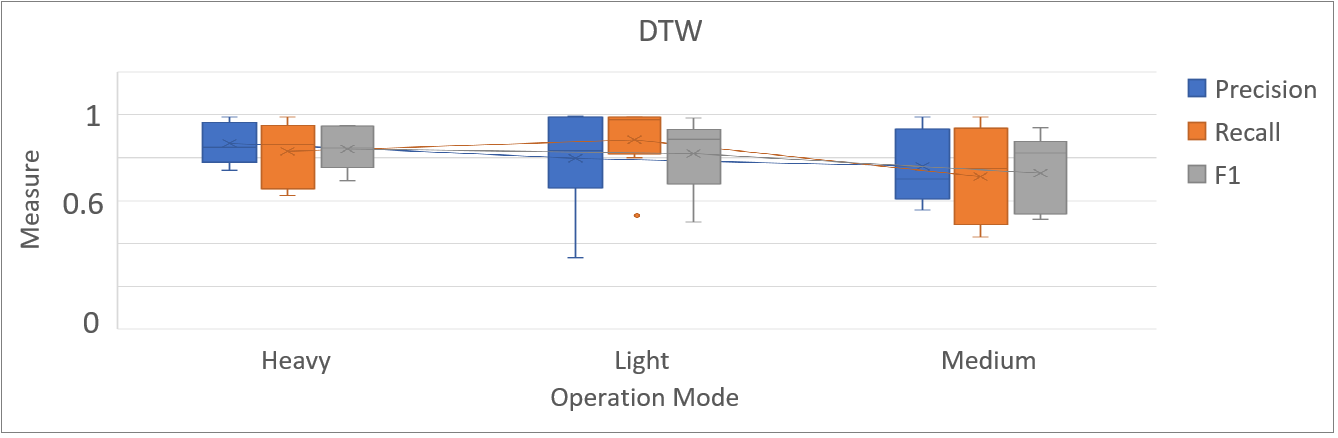}
		\caption{The precision, recall and F1-score for each operation mode using DTW. }
		\label{fig:DTWbox}
	\end{figure}

To test the performance of the DTW classifier, we generate power consumption data for three different households using the simulator.  These datasets are distinct in appliance usage in terms of $I_h$. We assume that for each appliance, there are three different levels of $I_h$: High, Medium, and Low.

We use three metrics to measure the performance of the classifiers we use. The metrics are the following: Precision, Recall, F1-score. Figure.\ref{fig:DTWbox} shows that for the DTW the metrics values are averaged around 82\% for the light and heavy operation modes. Otherwise, the metrics value is approximately 79\%.

\begin{figure}
	\centering
	\fbox{\includegraphics[trim={1.7cm 1.7cm 1.7cm 1.7cm },clip,width=.47\textwidth]{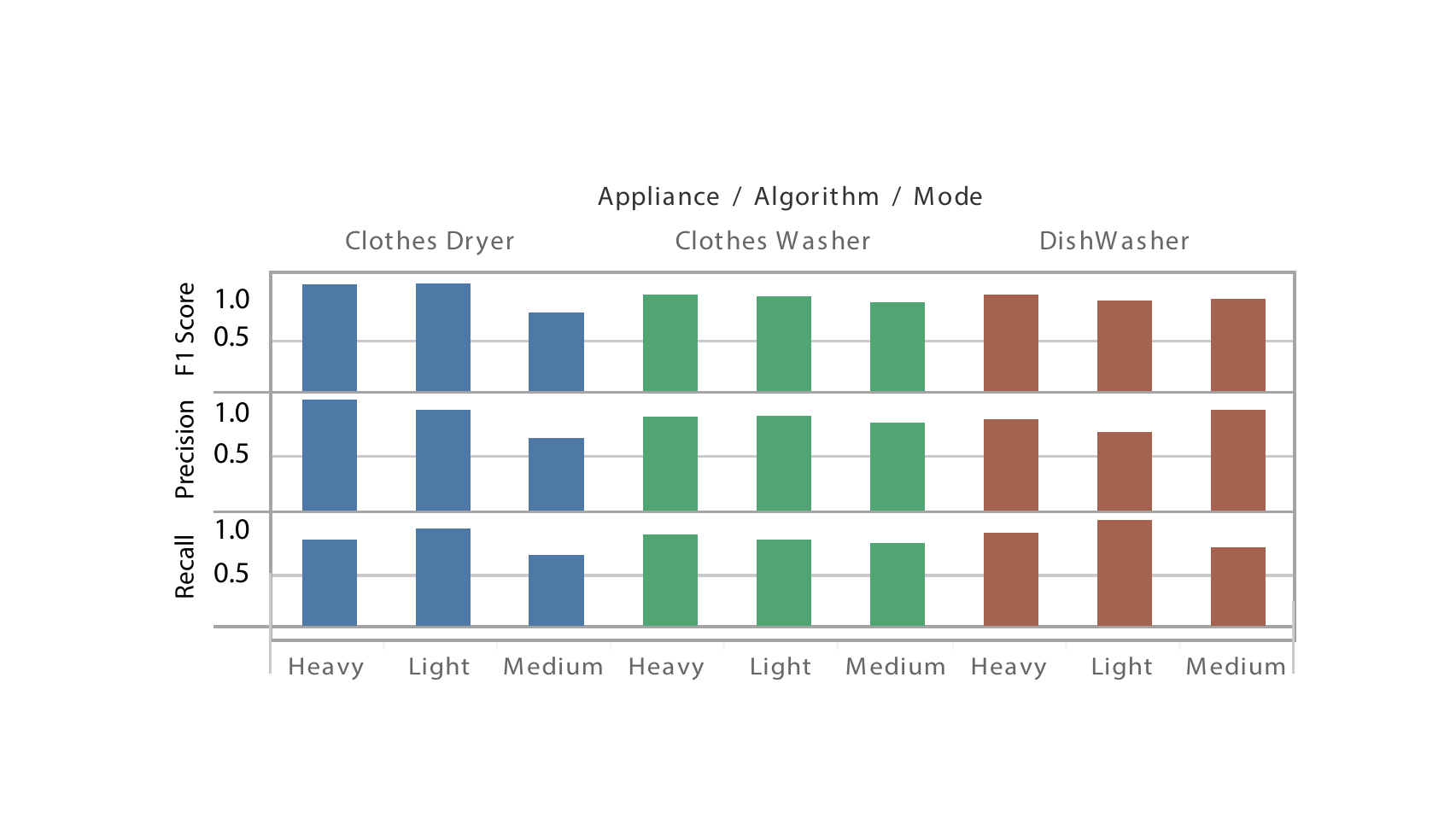}}
	\caption{The breakdown of the operation modes performance for each appliance using DTW and KNN algorithms.}
	\label{fig:evalAll2}
\end{figure}

The breakdown of the operation modes performance for each appliance  is shown in Figure.\ref{fig:evalAll2}. The chart is divided into three main lanes, each lane corresponds to an appliance. Within each lane, it shows the different metric values for DTW classification results over the operation modes. Generally, the chart shows higher performance for the light and heavy operation modes. In the clothes washer lane, there is a variation in the metrics values from operation mode to another. However, in the clothes dryer and the dishwasher lanes,it shows higher performance for heavy and light modes more noticeably in the clothes dryer.

%todo---------------------------------------------------------------------------------------------
\section{Conclusion And Future Work}
\label{conc}
Our work is focused on providing techniques built on top of residential power consumption for certain appliances. These techniques serve as the foundation that can built upon to better support DR. We analyzed PCD \cite{Makonin2017} by applying statistical analysis to find out statistical distributions about consumption for the households in the dataset. Furthermore we used some of the analysis in the literature \cite{Pipattanasomporn2014} to understand operation modes for appliances and extract their characteristics. A simulation engine then processes the statistical models resulted from the data analysis and generates power consumption data to simulate households use their appliances in different operation modes. This simulated data is used to test our models. An SUP detection algorithm is proposed using cross correlation between reference patterns and daily usage data to detect the activation times of appliances. These activation times are used by the recognition algorithm (DTW) to classify SUPs into their operation modes.
A future improvement to the current work is to utilize other versions of DTW such as AWrap \cite{Mueen2018} for sparse time series that is faster the original DTW. This provides opportunities to apply this work on online power consumption data streams collected by smart meters and sensors.

\bibliographystyle{plain}
\bibliography{bib}

\end{document}